%% file: chapter_csaas.tex
\documentclass[titlepage]{scrartcl}

\usepackage{type1cm}        
\usepackage{makeidx}         
\usepackage{graphicx}        
\usepackage{multicol}        
\usepackage[bottom]{footmisc}

\usepackage{newtxtext}       %
\usepackage[varvw]{newtxmath}       
\usepackage{wasysym}
\usepackage{array}
\newcolumntype{P}[1]{>{\centering\arraybackslash}p{#1}}

\usepackage[acronym]{glossaries}
\loadglsentries{acronyms}

\usepackage[colorinlistoftodos]{todonotes}
\usepackage{fancyvrb}
\usepackage{booktabs}
\usepackage{arydshln}
\usepackage[sorting=none]{biblatex}
\bibliography{bib}

\begin{document}
\title{Cybersecurity as a Service}
\author{John Morris$^*$ \and Stefan Tatschner$^*$ \and Michael P. Heinl \and Patrizia Heinl \and Thomas Newe \and Sven Plaga}
%
%
\def\thefootnote{*}\footnotetext{These authors contributed equally to this work.}\def\thefootnote{\arabic{footnote}}
\maketitle

\clearpage

Authors:

\begin{enumerate}
    \item John Morris$^*$; Department of Electronic and Computer Engineering, University of Limerick, Ireland; john.morris@ul.ie; ORCID: \url{https://orcid.org/0000-0003-2811-1055}
    \item Stefan Tatschner$^*$ Fraunhofer AISEC, Department Product Protection and Industrial Security, Germany; Department of Electronic and Computer Engineering, University of Limerick, Ireland; Confirm, the SFI Centre for Smart Manufacturing, Ireland;\\ stefan.tatschner@aisec.fraunhofer.de; ORCID: \url{https://orcid.org/0000-0002-2288-9010}
    \item Michael P. Heinl; Fraunhofer AISEC, Department Product Protection and Industrial Security, Germany; michael.heinl@aisec.fraunhofer.de; ORCID: \url{https://orcid.org/0000-0002-1094-4828}
    \item Patrizia Heinl; Technische Hochschule Ingolstadt, Germany; patrizia.heinl@thi.de; ORCID: \url{https://orcid.org/0009-0001-1594-2119};
    \item Thomas Newe; Department of Electronic and Computer Engineering, University of Limerick, Ireland; Confirm, the SFI Centre for Smart Manufacturing, Ireland; thomas.newe@ul.ie; ORCID: \url{https://orcid.org/0000-0002-3375-8200}
    \item Sven Plaga; Center for Intelligence and Security Studies (CISS), Germany; sven.plaga@unibw.de; ORCID: \url{https://orcid.org/0000-0002-1658-1140}
\end{enumerate}

\def\thefootnote{*}\footnotetext{These authors contributed equally to this work.}\def\thefootnote{\arabic{footnote}}



\clearpage

\abstract{With the increasing sophistication and sheer number of cyberattacks, more and more companies come to the conclusion that they have to strengthen their cybersecurity posture. At the same time, well-educated \gls{it} security personnel are scarce. Cybersecurity as a service~(CSaaS) is one possible solution to tackle this problem by outsourcing security functions to managed security service providers~(MSSP). This chapter gives an overview of common CSaaS functions and their providers. Moreover, it provides guidance especially for small- and medium-sized businesses, for asking the appropriate questions when it comes to the selection of a specific MSSP.}

\clearpage


\section{Introduction}
\label{sec:1}

\gls{csaas}, also sometimes referred to as \gls{secaas}~\cite{csa2016}, is the outsourcing of key \gls{it} security functions to an external specialist company or third-party. The concept of \gls{csaas} ultimately began back in 1987 with the  availability of the first antivirus product called VirusScan from McAfee \cite{ThePCinsider2023} where computer users paid to be protected from malware attacks. Roll on 30 years and as the malware has become more abundant and complex, the need for more protective services has increased in tandem. 
The initial uptake on this new breed of cybersecurity services with names like vulnerability assessment and \gls{ciso} s a service has been passive. One cause for this slow engagement is that many \glspl{ceo} believed investment in such services is an unnecessary expense. On the technical side, some \gls{it} Directors feel that their positions within the company structure is endangered and they are confident that they can do it better themselves, anyway. 
Particularly in the case where the outsourcing of key organisational security functions to outside contractors is concerned.

The recent increases in cyber-attacks of high-profile companies around the world~\cite{statista2023} and better cybersecurity education has altered this mindset in a positive way. Additionally, it has been proven that most organisations are still reactive when it comes to cybersecurity. They still believe that a malware attack will not happen to them: so why pay for cybersecurity? It is deemed too high a price for embracing the concept of precaution. However, when such deniers are stroke by a sudden malware attack, suffering untold data losses or paying ransoms to the cybercrime-as-a-service industry, these entities suffer greatly for their negligence. That is, if they are still even in business after the attack as currently over half of all small businesses close within six months of a malware attack~\cite{Cybercrime-2019}.

What is for certain though, is that the volume of malware attacks are set to increase and become more sophisticated, particularly with the advent of malware enhanced by \gls{ai} like DeepLocker~\cite{DeepLocker2018}, and few companies will have the expertise and resources to deal with this evolving cyber problem. Another point of note is that the malware attack surface is no longer confined to large networks of connected computers and servers, poorly written web interfaces, and email phishing attacks. The newer malware is targeting the entire \gls{ioe} landscape. From mobile phones to smart wearables, and resource-constrained \gls{iot} devices to cloud-based platforms. With such a large \gls{it} ecosystem to protect, it has become increasingly expensive for companies to train their \gls{it} staff to protect this attack surface or hire dedicated \gls{it} security staff. This is compounded by the fact that there is currently a worldwide shortage of \gls{it} security staff with current estimates at 3.4 million vacant positions~\cite{isc2022}. 

\gls{csaas} appears to be a step in the right direction to handling this growing threat landscape and allows companies to pick the \gls{it} security functions that they most need help with at a more affordable monthly rate. Simultaneously, not least due to the rising numbers of supply chain attacks, it is important that a provider is chosen who does not only offer an increase in security to its customers just from a technical viewpoint. To be able to protect sensitive customer data, a strong security ethos is also required on the provided services.
Over the course of this chapter, a more in-depth review of the most common \gls{it} security functions being offered by \gls{csaas} companies will be discussed. Also, a comparison of the main \gls{csaas} companies will be conducted. Finally, a checklist will be created for companies looking to choose a \gls{csaas} for themselves. 


\begin{figure}[ht]
    \centering
    \includegraphics[width=0.95\textwidth]{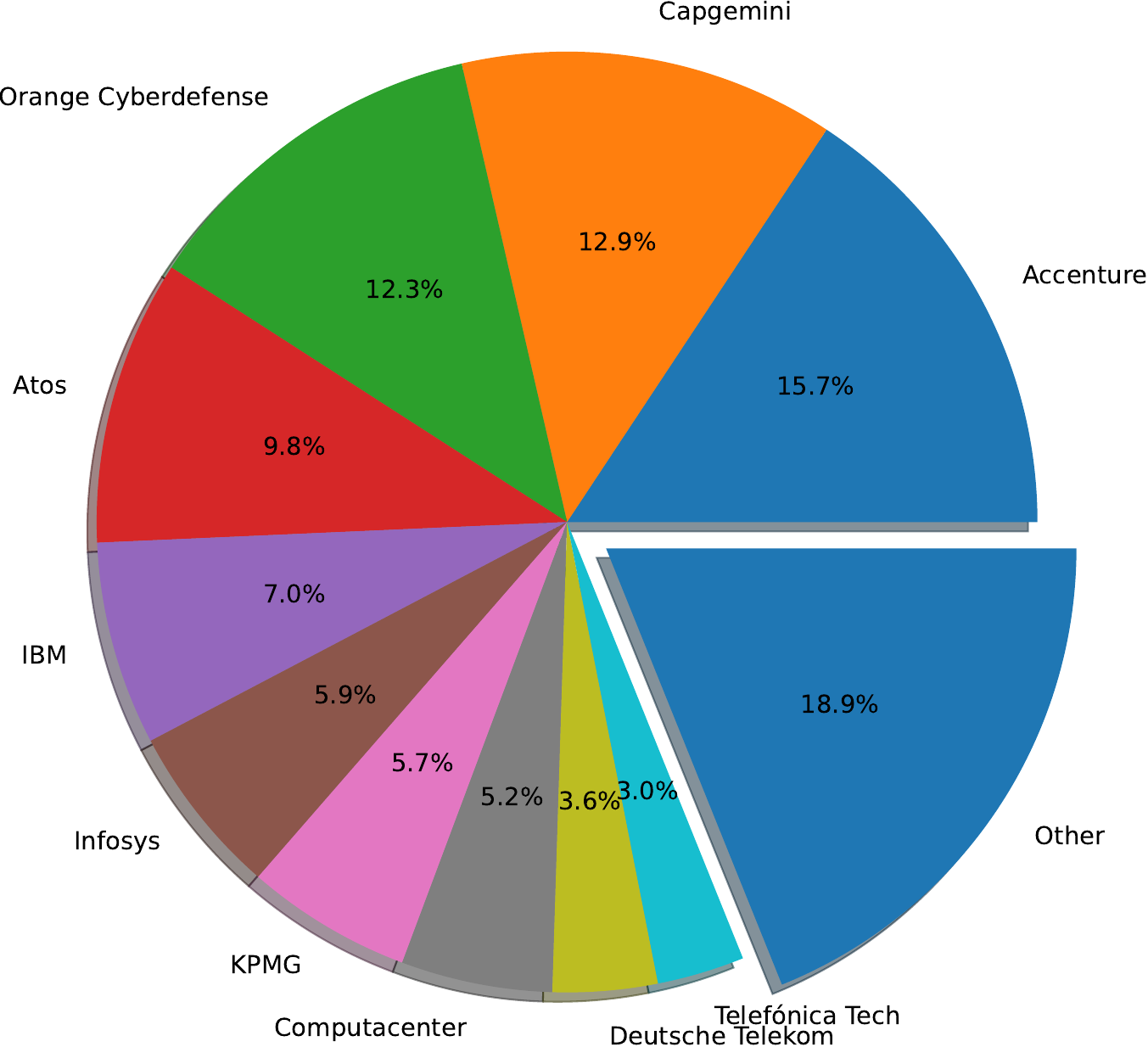}
    \caption{Managed \& Professional Security Services Market: Revenue Share of Top Participants, Europe, 2022; conducted by Frost \& Sullivan~\cite{frostSullivan2022}}
    \label{fig:revenue_share}
\end{figure}

The cybersecurity market has developed into one of the most profitable \gls{it} markets over the last decade \cite{statista-market-2023}. Consequently, a lot of new  \gls{it} companies specialised in cybersecurity were only founded in recent years or where existing \gls{it} companies launched dedicated cybersecurity divisions. According to the revenue study shown in Figure~\ref{fig:revenue_share}, the top ten companies in the managed and professional security services market in Europe are:
\begin{itemize}
    \item \textbf{Accenture} (\url{https://www.accenture.com}),
    \item \textbf{Capgemini} (\url{https://www.capgemini.com}),
    \item \textbf{Orange} (\url{https://orange.com}),
    \item \textbf{Cyberdefense} (\url{https://www.cyberdefensecompany.com}),
    \item \textbf{Atos} (\url{https://atos.net}),
    \item \textbf{IBM} (\url{https://www.ibm.com}),
    \item \textbf{Infosys} (\url{https://www.infosys.com}),
    \item \textbf{KPMG} (\url{https://www.kpmg.us}),
    \item \textbf{Computacenter} (\url{https://www.computacenter.com}),
    \item \textbf{Deutsche Telekom} (\url{https://www.telekom.com}), and
    \item \textbf{Telefónica Tech} (\url{https://www.telefonica.com}).
\end{itemize}

\gls{csaas} companies typically offer services in several forms, for instance subscription or payment for utilised services. In contrast, there are also variants where basic usage is free to use, but additions (e.g., 24/7 customer support, higher rate limits, or additional premium features) are charged. 

Outsourcing key \gls{it} security functions comes with benefits like cost cutting, a consistent and unified architecture, or better security expertise (by the \gls{csaas} company). On the other hand, implementing \gls{csaas} relies on sensible data being sent to the service provider which introduces multiple challenges requiring a well-designed architecture to avoid insecure applications. Consequently, companies offering \gls{csaas} \emph{must} maintain a good reputation in the marketplace and be trusted to stay relevant.
The importance of a good reputation for companies offering \gls{csaas} begs the question of decent selection. When looking to choose a \gls{csaas} company to engage with, what are the ten most common traits to look for?

\begin{enumerate}
    \item \textbf{How long is the entity in business?} \\
    The reputation is easier to spot when the entity is in business for a long time.
    In this case there may be online reviews, news articles, or similar material from third parties available.
    
    \item \textbf{What companies is the entity working with already?} \\
    Collaborating with big players in the same area of work can be a hint for a good and trusted reputation. Particularly, it these are long term customers. 

    \item \textbf{What range of services does the entity offer?} \\
    Offering few services could be a hint for a highly specialised entity offering high quality services. Are the specific services that are being sought, being offered by the entity?

    \item \textbf{What kind of service delivery model is employed? On-premise, remote, or both?} \\
    This trait is very specific to the relevant use case and the current security posture of the client company.
    On-premise means that dedicated resources and staff need to be provided, but a certain level of control is still ensured.

    \item \textbf{Is it a fully managed service or do internal \gls{it} resources have to be dedicated to delivering the services?}\\
    This depends on whether the client company has internal staff with the requisite skills and time to manage the security requirements of the company. Fully managed is designed for client companies with little or no internal security personnel or systems.

    \item \textbf{What type of pricing model is offered? Fixed monthly, annually, per employee or device?}\\
    This will depend on the type of security service being offered. Security training is typically charged by employee whereas penetration testing and cyber insurance can be charged monthly or annually. 

    \item \textbf{What is the skillset and qualifications of the staff?} \\
    Are the staff certified or doing public speeches at conferences in their area of work? Is their training relevant and kept up to date? Where are the gaps in the security staff skills that need to be filled by an external security company?
    
    \item \textbf{Has the entity published any articles or does the entity take part in any blogs or forums in the areas of cybersecurity?} \\
    This is a big indication of a security company that is highly skilled and extremely competent. It also means that they are keeping up to date with the latest security threats and trends.

    \item \textbf{Does the entity provide a trial period or proof of concept?}\\
    This can be helpful in deciding if a particular security company or tools is compatible with the needs of a client company. A proof of concept can provide a try before you buy type scenario to help key decision makers in the approval process. 

    \item \textbf{Is the entity certified?} \\
    Certifications help ensuring that at least a minimum level of security. It also gives the client company a comfort in knowing that the entity has the requisite security qualifications to complete the security services being offered.
\end{enumerate}

This book chapter contributes a list of ten most common traits to look for when choosing a \gls{csaas} company.
In addition to these traits, common \gls{csaas} functions are researched and are related with high revenue companies.
Furthermore, an overview over the current market share of professional \gls{csaas} providers with a comparison about the offered services is given.

\section{\gls{csaas} Functions }
\label{sec:2}

The number of different cybersecurity services offered by these companies are substantial, especially when specialised use cases are included. However, the Cloud Security Alliance has published an overview \cite{csa2016} where a categorisation of cybersecurity services was carried out. The provided categorisation was enhanced by additional services based on our practical knowledge and logical reasoning. The identified key services are described in the following sections.

\subsection{Security Personnel as a Service}
\gls{ciso} as a Service or Virtual \gls{ciso} is the outsourcing of the Chief Information Security Officer role within an organisation. This resource can work onsite within a particular organisation or work remotely; reporting directly to the C-Level Group which is key for decision making. They can work independently or as the head of a security team, work for a fixed contract period or month-to-month. Their duties include:  

\begin{itemize}
    \item Full review of an organisation’s security position. 

    \item Recommend best practise hardware, software and security changes. This can also include purchases. 

    \item Interview, vet and hire new security staff 

    \item Train internal security team. 

    \item Generate penetration testing report. 

    \item File NIST 800 security reports where required.  
\end{itemize}

This role is more suited to mid to large sized companies where the budget for a permanent \gls{ciso} role is currently not available or as a try before you buy type scenario. A main constraint of this approach is the often steep learning curve for the contractor in terms of corporate knowledge, cultural norms and company politics. However, this last point can also be an advantage as the contract \gls{ciso} is not affected by internal conflicts or job security.  

Additional security roles that can be outsourced include a data protection officer, compliance and risk officer, forensic analyst, security trainer, penetration tester and security helpdesk personnel.  

\subsection{Cyberawareness Training}
\begin{figure}[h]
\centering
\frame{\includegraphics[width=0.9\textwidth]{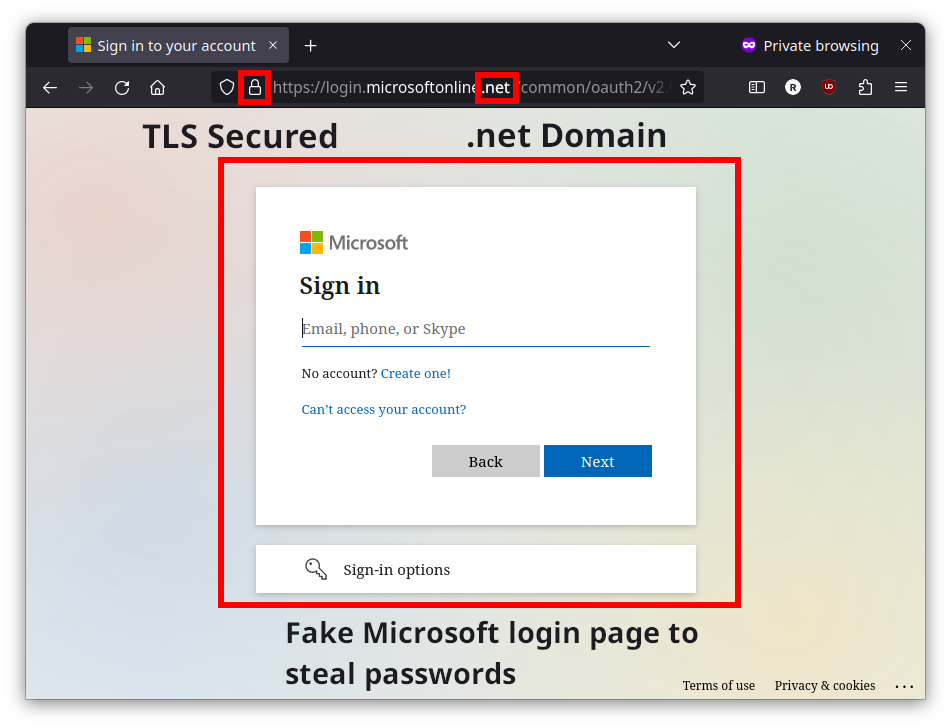}}
\caption{Screenshot of a phishing website for a Microsoft login \cite{fishing-site}.}
\label{fig:websitephishing}       
\end{figure}

Cyberawareness or malware threat detection training involves the systematic education of company employees in how to correctly identify malware threats, since 95\% \cite{IBM-Cybersecurity-2022} of current company malware breaches are caused by human error. The format of the training is usually a step-by-step guide containing videos and a series of items to identify afterwards, to reinforce the training. The training usually finishes with a quiz of all the topics discussed in the session with a completion certificate produced for a passing grade. The most popular cyberawareness training programmes concentrates on email phishing and social engineering attacks. In other words, training employees to think before clicking on that web link and entering their login credentials into a fake website like in figure \ref{fig:websitephishing}.

The training normally lasts around 30 to 40 minutes with some like the Kevin Mitnick inspired KnowBe4 email phishing offering lasting 50 minutes. The cyberawareness training is then reinforced further with weekly mock phishing attacks being sent out to all employees. Training should be retaken by employees at least once a year to keep abreast of new types of malware attacks. The training is offered as a managed service that typically reports to the \gls{hr} department rather than \gls{it}. The main types of cyberawareness training sessions include: 

\begin{itemize}

    \item Phishing, Smishing and vishing attacks. 

    \item Remote work training. 

    \item \gls{gdpr} training. 

    \item Foreign travel dos and don’ts. 

    \item Intellectual or physical company property training.
         
\end{itemize}

The cyberawareness training can also be customised with corporate branding and content to make it more realistic to the employees (e.g., actual company emails) and assist in the process of turning them into human firewalls.

\subsection{Vulnerability Assessment}

A vulnerability assessment involves the systematic identification, measurement and categorisation of weaknesses within an organisation’s systems. These weaknesses can take the following forms:

\begin{itemize}

    \item Unpatched and unencrypted servers and/or computers. 

    \item Poorly setup firewall with open rules and port access. 

    \item Remote access vulnerabilities. 

    \item Software and application unauthorised access. 

    \item Lack of document lock storage cabinets or shredding facilities. 

    \item Poor website design with limited security and/or no \acrshort{tls} encryption. 

    \item Faulty door locks or doors left open  

    \item Weak or no password policies. 

    \item No document or data audit process. 

    \item Weak or no Wireless Access Point security. 

    \item Employees susceptible to social engineering attacks. 
    
\end{itemize}

Typically, an off-the-shelf vulnerability scanner is used to identify weaknesses within an organisation. Current scanners can identify over 100K separate system vulnerabilities in as little as an hour; depending on the system size and complexity \cite{holmsecurity2023}. In the absence of in-house security personnel to conduct the assessment, it can be conducted using external security personnel. However, to complete the assessment properly, all systems will need to be scanned from inside the organisation as well as from the outside. Once the assessment is complete, a detailed vulnerability report is created based on the weaknesses listed above. The vulnerabilities are classified by severity and frequency. A separate executive report is normally produced for the key decision makers with less detail and more emphasis on the risks and financial impact to the organisation. 

\subsection{Periodic Penetration Testing}

Periodic Penetration Test is an authorised simulated cyberattack on a computer system, performed on a regular basis to evaluate the security of the system. Its objective is to identify vulnerabilities that could otherwise be used by malicious actors to abuse the computer system. A Penetration Test needs to be performed by a technical domain expert who can use similar techniques as those used by attackers. 

Penetration Testing is a demanding task, and the following challenges apply:
\begin{enumerate}

    \item \textbf{Staying up to date} with the current state of the art from a technical standpoint. The \gls{it} sector is developing at a very fast pace and a penetration tester must be capable of all the current and relevant technologies when conducting an effective test. 

    \item \textbf{Scope:} Defining the scope of testing is a challenging task. On the one hand, a scope that is too narrow might not yield useful results. On the other hand, too broad a scope could be unfeasible from a management perspective. 

    \item \textbf{Realistic Attack Scenarios} are considerable for a penetration test, since a highly academic attack scenario could indeed yield results. However, these results are at risk of not being relevant for the desired use case of the product. 

    \item \textbf{Limited Access:} The integration of cybersecurity in the development process (i.e. security by design) is desired, since technical design decisions often have an impact on the security of a system. However, penetration testing during development can be restricted, since parts of the system might not be implemented yet. 

    \item \textbf{Reproducing Issues:} Reproducing findings needs the careful documentation of all involved working steps and parameters of the test environment. Monitoring every relevant parameter in a penetration test is a difficult task, since all included parameters might not be known by the penetration tester at the offset. 

    \item \textbf{Time Constraints:} Penetration testing is a complex task including creative components where good findings do not strongly correlate to the amount of time being spent on a test. However, budgeting in the first place can limit the effectiveness of penetration testing, since it limits the creativity of the tester. 

    \item \textbf{Collaboration and Integration} with the development team is required for the feedback loop to integrate any findings improving the actual product. 

    \item \textbf{Skills:} Finally, the skillset of the penetration tester must be accurate for the relevant architecture and used technology.  
    
\end{enumerate}

Security by Design is becoming more and more important in the design process of software products. Companies are beginning to integrate Secure Software Engineering into the relevant value chains \cite{9328095}. Periodic Penetration Testing is a good option for evaluating that the designed software architecture is secure and that included security measures serve their purpose. However, in order to be effective, it requires careful planning and implementation.

\subsection{Email Security}

E-mail security is a critical component of an organisation’s communication. Due to its legacy, e-mail suffers from many design issues related to security. For instance the \emph{content} of an e-mail is usually only secured from the e-mail client to the e-mail server rather than being end-to-end secure. E-mail was designed at a time when the internet was mainly an academic tool and thus end-end-security was not relevant. However, the success of e-mail especially in a corporate context might be a result of this simplicity. 

There are several key technologies available which are implemented by default by the common big e-mail service providers. Since e-mail does not provide any of these technologies by default, they were added on top, for example, adding metadata via e-mail headers. 

\begin{enumerate}

    \item \textbf{Encryption:} A procedure of converting plain text into a so-called cipher text, which can only be decrypted with a specific key. Encryption implements the protective goal of confidentiality both at transit and at rest. Most commonly used state of the art technologies are \acrfull{smime} or \acrfull{pgp}.

    \item \textbf{Digital Signatures:} Digital Signatures are used to verify the authenticity and integrity of messages by using special metadata which is attached to a message. In other words, these signatures can be used to verify that the message has not been tampered with during transit and that it was sent by the claimed sender. Most commonly used state of the art technologies are \acrfull{smime} or \acrfull{pgp}.

    \item \textbf{Spam Filters:} Filters which use sophisticated techniques to block unwanted messages. 

    \item \textbf{Anti Malware Solutions:} Use signature-based detection, heuristics, or machine learning to identify and block messages that contain malware, such as viruses, Trojans, or spyware. 

    \item \textbf{\gls{spf}:} A protocol that allows organisations to specify which mail servers are authorised to send e-mails on their behalf. 

    \item \textbf{\gls{dkim}:} A protocol that allows organisations to digitally sign e-mail messages on the server side to verify the authenticity and integrity of the message. 

    \item \textbf{\gls{dmarc}} A protocol that allows organisations to protect their domains from unauthorised use, such as phishing and e-mail spoofing. \gls{dmarc} allows organisations to publish policies that specify how recipient mail servers should handle e-mails that fail \gls{spf} and \gls{dkim} authentication. 

    \item \textbf{\gls{arc}:} A protocol that provides a chain of authentication results for an e-mail message, starting from the original sending mail server to the recipient's mail server. 

    \item \textbf{\gls{tls}:} A protocol that is used to provide communications security over a computer network. Due to its current widespread use in instant messaging, file transfers and web traffic, \gls{tls} has become a basic technology for secure internet today. 
\end{enumerate}

The following Listing~1 shows the added header fields and the structure change of an e-mail with ARC, DMARC, DKIM, SMIME, and SPF in place.
Items in \textbf{bold} face are added by these extensions.

\begin{Verbatim}[fontsize=\scriptsize,frame=single,label={Listing 1: Structure of an e-mail with ARC, DMARC, DKIM, SMIME, and SPF},labelposition=bottomline,commandchars=\\\{\}]
From: sender_email_address
To: recipient_email_address
Subject: email_subject
\textbf{MIME-Version: 1.0}
\textbf{ARC-Seal: arc_seal_value}
\textbf{ARC-Message-Signature: arc_signature_value}
\textbf{DKIM-Signature: dkim_signature_value}
\textbf{DMARC-Record: dmarc_record_value}
\textbf{Received-SPF: pass (sender_ip_addr: domain_of_sender designated_server_ip_addr permitted)}
\textbf{Authentication-Results: domain_name;}
\textbf{    spf=pass smtp.mailfrom=sender_email_address;}
\textbf{    dkim=pass header.i=@domain_name;}
\textbf{Content-Type: application/pkcs7-mime; smime-type=enveloped-data; name=smime.p7m}
\textbf{Content-Disposition: attachment; filename=smime.p7m}
\textbf{Content-Transfer-Encoding: base64}

\textbf{base64_encoded_SMIME_message_body}

\end{Verbatim}

Due to this added complexity on top of the basic e-mail design, running a secure e-mail service is relatively cumbersome. Especially as a violated or missing protocol could impair successful delivery of e-mails. Consequently, there are several companies that are specialised in providing secure e-mail services. Well known free e-mail providers utilising most of the mentioned key technologies are Google with its GMail\footnote{https://gmail.com} service and Microsoft with Exchange\footnote{https://outlook.live.com}.

\subsection{Identity and Access Management}

\gls{iam} is a basic requirement of every effective security program in order to protect data, applications, and other assets. To be able to technically enforce it, i.e., only authorise legitimate requests, users must be reliably authenticated. This is usually done leveraging digital identities, e.g., usernames, which are linked to a person’s actual identity. Typical standards used in this context are OAuth~\cite{rfc6749}, OpenID~\cite{OpenIDCore}, and \gls{saml}~\cite{rfc7522}. Establishing and managing these digital identities seems to be a straightforward task but can become very complex once the number of employees and other stakeholders of an organisation increases. 

Therefore, \gls{iam} providers do not only offer the corresponding technologies but also best practices in the form of pre-defined processes and concepts. Typical functionalities offered by \gls{iam} providers include but are not limited to: 

\begin{itemize}
    
    \item Initial registration of users. 

    \item Assignment of roles and privileges. 

    \item Creation, provision, and management of credentials. 

    \item Centralised management of identities, roles, and privileges. 

    \item Centralised authentication and authorisation of users 

    \item Provision of means for \gls{mfa} 

    \item Support of interfaces for \gls{sso} services 
    
\end{itemize}

Accounts with a very high level of privileges, e.g., administrators or superusers, are a popular target of threat actors and prone to insider risk. They should therefore be additionally protected leveraging \gls{pam}. 

\subsection{Cyber Insurance}
In the last few years, the frequency and impact of cyber incidents against companies worldwide continued to increase steadily~\cite{embroker2023}.  While some industry segments were hit less frequently than others~\cite{technews2023}, there is no guarantee for anyone to be spared to move into the focus of threat actors. Hence, no matter how much money a firm spends on its security program, or which technical prevention controls it implements, there is a residual risk of being hit by a cyber-attack that might lead to reputational and/or financial loss for the victim. 

The purpose of Cyber Insurance is to step in if an insured victim experiences such a reputational or financial loss arising out of a covered cyber incident. Coverages that are generally offered by insurance companies include: 

\begin{itemize}
    \item     First party damages (i.e., losses directly occurred to the policyholder) covering own costs (e.g., business interruption costs, incident response and forensics expenses, the launch of public relation campaigns, installation of call centres to inform customers).
    \item     Third party liability (e.g., claims made against the policyholder by a third party) covering costs to indemnify the claimants for a loss and the expenses of defending lawsuits associated with it. In many cases, these losses arise from the failure of an organisation to appropriately protect third parties’ data from being breached or compromised through a cyber incident.  
\end{itemize}

Additionally, many insurance carriers offer further services to their customers such as establishing connections to forensic and incident response firms as well as consultancy services. This is beneficial for both, the insurance carriers and the insureds, as both are interested in quick recovery after an incident to reduce costs. 

While the process for a company getting cyber insurance certainly can differ, there are some steps each carrier performs before offering a binding quote for cyber coverage: 

\begin{enumerate}
    \item Assessment of cyber exposure based on industry, company size, and business model.

    \item Evaluation of security protection level by on-site visits, conversations, questionnaires, and/or cyber risk scanning and analytics tools. 

    \item Legal wording of cover elements and exclusions.

    \item Actuarial calculation of potential losses, maximum capacity, and corresponding premium. 
\end{enumerate}

With the recent surge of cyber incidents, insurance companies started to be more selective on offering cyber insurance. Companies need to fulfil minimum security standards defined by each carrier. In addition to that, insurers need to protect themselves from large scale events which can hit multiple clients at once, so-called accumulation risks. Scenarios which are under discussion and currently excluded by most carriers are cyber incidents which arise out of any kind of cyber war (whether declared or not) and the outage of external networks, such as the internet or electricity supply. 

\subsection{Incident Response}

\begin{figure}[h]
\centering
\includegraphics[width=1\textwidth]{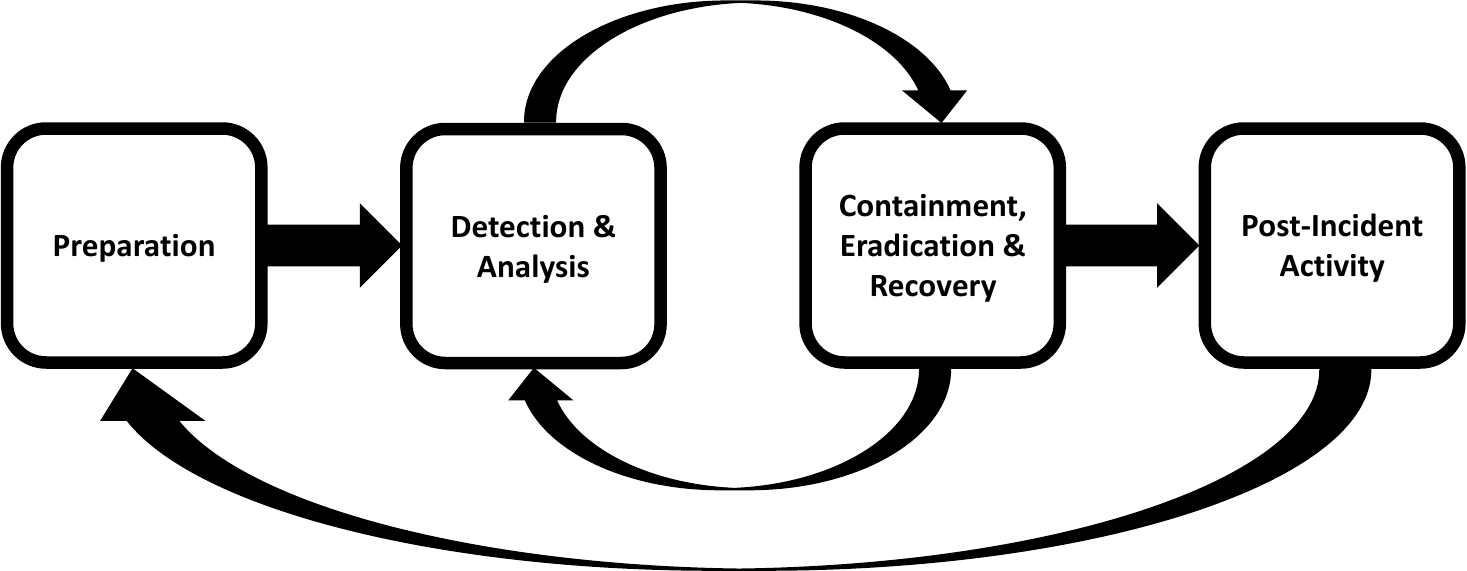}
\caption{Incident response life cycle according to \gls{nist} SP 800-61 Rev. 2 \cite{NISTSP800-61Rev2}.}
\label{fig:nistirlc}       
\end{figure}

There is a saying that companies should not ask themselves if they are vulnerable to a security incident but only when and to which extent this incident may occur. Keeping that in mind, it is important to be prepared for the moment in which such an incident happens. Therefore, \gls{ir} services should not only provide support during an incident. According to the \gls{nist}, the incident response life cycle encompasses a total of four phases as shown in Figure~\ref{fig:nistirlc}:
\begin{enumerate}
    \item Preparation.
    \item Detection and Analysis.
    \item Containment, Eradication, and Recovery.
    \item Post-Incident Activity.
\end{enumerate}

Ideally, an \gls{ir} service covers all of these phases. This makes rapid response much more likely, as information from all phases is directly available during the actual \gls{ir} and does not have to be shared cross-organisationally among different service providers, which would cost valuable time.

Before the actual incident, incident response services encompass consultation on technology enabling the customer to detect and contain incidents, e.g., solutions for \gls{siem} and Endpoint Detection and Response (EDR). Furthermore, one of their technological focuses is on configuring the customers’ infrastructure not only securely but in a way that retains and protects information which is valuable for incident handling and investigation, e.g., read-only backups and audit logs. Apart from these technological measures, \gls{ir} also encompasses preparation on an organisational and human level, including the preparation of customised response plans and playbooks as well as regularly putting their content into practice through tabletop exercises. Ideally, these tabletop exercises are as inclusive as possible, involving not only representatives from \gls{it} (security) but also from operations, legal, human resources, public relations etc. 

For the case where a potential incident has been detected, \gls{ir} services ideally offer an emergency hotline which can be consulted 24/7 in order to provide support during the process of triage and first response. Once it is confirmed that the initial alarm has not been a false positive, \gls{ir} services begin with evidence collection and root cause analysis. In order to be prepared for potential court cases and to support law enforcement, it is paramount to document the analysis as thoroughly as possible and maintain the chain of custody during forensics.

When affected parts of systems and networks are identified, an appropriate containment strategy, such as powering them off or disconnecting them from other parts of the network, has to be chosen. The choice heavily depends on the pursued, sometimes conflicting objectives besides the actual containment, e.g., preserving evidence even in non-persistent memory or stopping a ransomware attack from continuing to encrypt data. Once the threat is contained, it has to be eradicated, e.g., by wiping malware, mitigating vulnerabilities, and disabling compromised accounts. After that, recovery can take place, e.g., by resetting passwords and restoring systems.

As indicated in Figure~\ref{fig:nistirlc}, the described phases are not strictly linear but rather part of an iterative, recurring process. Depending on the organisational and technological environment of the individual incident, \gls{ir} engagements can happen on premise, remotely, or in a mixed mode, depending on the phase.

\subsection{Business Continuity / Disaster Recovery Planning}

The planning of \gls{ir} and \gls{bcdr} are closely related. However, the scope of \gls{bcdr} goes beyond potential business interruptions caused by security incidents and does primarily focus on the continuity and recovery of the core business, i.e., keeping critical processes running independently from the environment or restore them as quick as possible, respectively. Since these core processes change over time, \gls{bcdr} also must dynamically adapt and is therefore not a task to do once but a continuous process which can be managed systematically according to ISO 22301. Just as \gls{ir}, \gls{bcdr} is a highly interdisciplinary process involving various stakeholder groups to discuss and define a desirable yet realistic \gls{rto}, \gls{rpo}, as well as the corresponding measures. \gls{bcdr} as a Service can include the organizational part of moderation, consolidation, and documentation of these stakeholders’ requirements in the form of a \gls{bcdr} plan but also what is called Recovery as a Service, meaning backup and restore solutions hosted in the cloud. 

\subsection{Security Information and Event Management}
As previously mentioned, \gls{siem} can be very helpful when it comes to the detection and investigation of security incidents. Besides the pure aggregation of potentially security-related information, e.g., log files or real-time network data, from a variety of sources, it can also offer continuous monitoring and correlation to automatically (e.g., by anomaly detection) or semi-automatically (e.g., by pre-configured use cases) detect suspicious activities. Additional factors to be considered are intuitive user interfaces and flexible support of formats and protocols to include data from as many nodes as possible, as well as the scalability to be able to serve the dynamic landscape of a growing business. Apart from the option to deploy and use it on premise, it can also be deployed in the cloud and observed by well-trained analysts of the provider, ideally working in shifts to provide 24/7 coverage. This comes with the advantage that security alerts can be analysed directly when they happen, i.e., without long delays after business hours or on weekends. 

\subsection{System Patching and Updates}
With the disclosure of software vulnerabilities, vendors are required to correct them as fast as possible, since they might be discovered and exploited by attackers to gain access to a computer system. Reacting as fast as possible to disclosed vulnerabilities is commonly called patching, since it is critical to preempt attackers. Good historical examples where software updates were mission critical are Heartbleed\footnote{https://cve.mitre.org/cgi-bin/cvename.cgi?name=CVE-2014-0160}, Triple-Seven\footnote{https://cve.mitre.org/cgi-bin/cvename.cgi?name=CVE-2016-0777}, Shellshock\footnote{https://cve.mitre.org/cgi-bin/cvename.cgi?name=CVE-2014-6271}, and EternalBlue\footnote{https://cve.mitre.org/cgi-bin/cvename.cgi?name=CVE-2017-0143}. What these vulnerabilities have in common is a large and possibly fatal impact on the attacked \gls{it} infrastructure:
\begin{itemize}
    \item They can be easily discovered by an attacker.
    \item They are easily exploitable (usually few lines of e.g. Python code).
    \item They have a fatal impact, for instance \gls{rce} or sensitive information leaks.
\end{itemize}

Fortunately, software updates for such kinds of critical vulnerabilities usually are available very quickly. For instance, patches for the famous Heartbleed vulnerability were available even before it was privately disclosed to the development team. Seven days after the disclosure an official release of the affected software was available\footnote{https://www.smh.com.au/technology/heartbleed-disclosure-timeline-who-knew-what-and-when-20140414-zqurk.html}. At the time of disclosure there were a round 300k vulnerable servers online. It is surprising that six years later there were still 200k vulnerable servers online\footnote{https://isc.sans.edu/diary/26798}.

These examples show the necessity of keeping up with evolving threats. Therefore cyber security systems need to track the current state of the art of available countermeasures. For instance, software modules that process untrusted data are one of the most critical parts to protect, as they are directly accessible by attackers. Operating systems provide mechanisms offering basic protection which in general limit the attack surface. In order to benefit from such cautionary measures regular security updates and reviews are desired. 

Software updates in production are rolled out via well-established update mechanisms. In \gls{foss} environments packet management systems, such as \texttt{apt}, \texttt{dnf}, or \texttt{pacman} are common. Usually, there are different update tracks including stable updates (i.e., stability and security updates) or bleeding edge (i.e., new features are deployed as fast as possible). In non-\gls{foss} environments there might be proprietary solutions with similar semantics. Careful reviews of the used software repositories are required when building products or infrastructures relying on these updates. \gls{csaas} companies ensure that maintained components or services stay up-to-date and are not affected by known vulnerabilities.

\subsection{Security Standards Compliance}
With the rising number of networked devices and digitisation of most parts of our lives in the context of the Internet of Everything, the number of security-related regulations and industry-specific standards which need to be considered continuously increases. Examples include: 

\begin{itemize}

    \item General Data Protection Regulation (GDPR), 

    \item ISO/IEC 27001 Information Security, 

    \item ISA/IEC 62443 Cybersecurity for Operational Technology, 

    \item ISO/SAE 21434 Road Vehicles -- Cybersecurity Engineering,  

    \item NIST 800-171 Security controls and processes for data protection,
    \item Cybersecurity Maturity Model Certification (CMMC) program,
    \item the European Cyber Resilience Act, and more to come.

\end{itemize}

Auditing the compliance with the requirements defined in these documents requires subject matter expertise and can be time-consuming. Therefore, it is often outsourced. With more and more services in the cloud, there are also approaches to check the compliance with specific requirements fully automated~\cite{stephanow2017clouditor}. 

\section{Future of \gls{csaas}}
Future \gls{csaas} offerings will potentially have to consider different currently ongoing trends in the security landscape. For example, there is the already mentioned threat of \gls{ai}-enhanced malware. However, \gls{ai} also poses other security threats to companies and public organisations, e.g., in the form of deep fakes or ChatGPT-generated spear-phishing campaigns. Considering the ease of use of tools like ChatGPT, tailored spear-phishing could have the potential to supersede normal Spam not only in terms of quality but also in numbers. Another trend is the increasing number of supply chain attacks~\cite{enisaThreatLandscapeSCA2021}. This may lead to an increased demand for \glspl{zta}, especially towards previously trusted third parties, which can be potential starting points for the mentioned supply chain attacks, as well as for enhanced protection of customer data needed to deliver specific managed security services, e.g., \gls{siem}. Moreover, existing trust relationships, for example, towards critical information infrastructures such as \glspl{ca}, have to be reconsidered and enhanced control mechanisms need to be established~\cite{mercat2019}. Eventually, the rise of quantum computers may not directly lead to new types of services. However, it will definitely have an impact on existing services. They will have to timely adapt to the new post-quantum algorithms once they are finally standardised by \gls{nist} to ensure future-proof security is also protecting against \textit{store now, decrypt later} type of threat scenarios.

\section{Findings and Suggestions}

The fact that 95\% of all company malware breaches are caused by human error \cite{IBM-Cybersecurity-2022}, has precipitated in the volume of companies currently adopting cyberawareness training programs to increase by 15\% year on year to date and the cyber awareness training market to reach a predicted \$10 billion annually by 2027 \cite{Awareness-Market-2023}. Additionally, the number of companies opting to pay for cyber insurance has risen steadily over the last three years partly due to a large number of high profile attacks during this time frame and the war in Ukraine. However, the uptake has now started to level off mainly due to the estimated 83\% hike in cyber insurance premiums over the last 12 months and the purchasing of better \gls{it} security equipment (e.g. next-gen firewalls and business continuity solutions) \cite{Insurance-Market-2023}. As working from home, either partly or totally, has become more mainstream for employees around the world, companies have had to look at new ways to protect their employees and intellectual property from malware attacks. As company \gls{it} staff cannot effectively protect all of these new remote working location, decision makers are opting for \gls{csaas} companies to assist with this large threat canvas. This new working model bodes well for the future growth of the \gls{it} security services industry. Finally, the new elephant in the room, from a security threat perspective, is the mobile phone. These ultra portable computers can now handle most of the day-to-day employee tasks like answering email, attending meetings, workflow approvals to reading and writing company documents. Most companies still overlook the security threat that mobile phones pose. They are finally taking action by installing anti-malware protection on these devices, allowing them access to guest wireless networks only and banning them from company meetings. 

\section{Conclusion}
It is important to mention that the protection demand of a specific organisation can be highly individual depending on factors, such as the sectors they are doing business in and the type of data they manage. The list of security services therefore only covers a selection of services which are most likely to be relevant for the majority of companies. When deciding which protection needs are applicable for an individual organisation, it is recommended to include representatives of the organisation’s stakeholders and utilise independent advice from external specialists, where needed. Companies employing connected manufacturing processes in the context of Industry 4.0, for example, might have an increased demand for monitoring focusing particularly on \gls{ics} or \gls{ot} which implies factors like safety and therefore another kind of security goal prioritisation. Explaining such sector-specific demands is not within the scope of this chapter. 

\begin{table}[h]
\caption{Mapping \gls{csaas} to top ten professional security providers according to Frost \& Sullivan \cite{frostSullivan2022}.}
\label{tab:mapCSaaStoProviders}       
%
%
\tiny
\begin{tabular}{p{2.5cm}P{0.8cm}P{0.8cm}P{0.8cm}P{0.8cm}P{0.8cm}P{0.8cm}P{0.8cm}P{0.8cm}P{0.8cm}P{0.8cm}}
\toprule
 & Accenture & Capgemini & Orange Cyber Defense & Atos & IBM & Infosys & KPMG & Computa- center & Deutsche Telekom & Telefonica Tech\\
\midrule
Security Personnel as a Service & \Circle  & \Circle & \CIRCLE & \CIRCLE & \CIRCLE & \Circle & \CIRCLE & \Circle & \LEFTcircle & \Circle \\
\hdashline
Cyberawareness Training  & \Circle & \CIRCLE & \CIRCLE & \Circle & \CIRCLE &  \CIRCLE & \CIRCLE&\Circle & \CIRCLE & \Circle \\
\hdashline
Vulnerability Assessment  & \CIRCLE &  \Circle & \CIRCLE & \CIRCLE & \CIRCLE & \CIRCLE & \CIRCLE & \CIRCLE & \CIRCLE &  \CIRCLE\\
\hdashline
Periodic Penetration Testing  & \CIRCLE &  \CIRCLE & \CIRCLE & \CIRCLE & \CIRCLE & \CIRCLE & \CIRCLE & \CIRCLE & \CIRCLE &  \CIRCLE\\
\hdashline
Email Security  & \Circle  &  \Circle & \CIRCLE & \CIRCLE & \CIRCLE & \CIRCLE & \Circle & \CIRCLE & \CIRCLE & \CIRCLE \\
\hdashline
\gls{iam} & \CIRCLE &  \CIRCLE & \CIRCLE & \CIRCLE & \CIRCLE & \CIRCLE & \CIRCLE & \CIRCLE & \CIRCLE &  \CIRCLE\\
\hdashline
Cyber Insurance  & \Circle & \Circle  & \Circle & \Circle & \Circle & \Circle & \Circle & \Circle & \LEFTcircle &  \Circle \\
\hdashline
Incident Response & \CIRCLE &  \CIRCLE & \CIRCLE & \CIRCLE & \CIRCLE & \CIRCLE & \CIRCLE & \CIRCLE & \CIRCLE &  \CIRCLE\\
\hdashline
\gls{bcdr} & \CIRCLE &  \CIRCLE & \CIRCLE & \CIRCLE & \CIRCLE & \CIRCLE & \CIRCLE & \CIRCLE & \CIRCLE &  \Circle \\
\hdashline
\gls{siem} & \CIRCLE &  \CIRCLE & \CIRCLE & \CIRCLE & \CIRCLE & \CIRCLE & \CIRCLE & \CIRCLE & \CIRCLE &  \CIRCLE\\
\hdashline
System Patching and Updates & \CIRCLE & \LEFTcircle & \Circle & \LEFTcircle & \Circle & \LEFTcircle & \Circle  & \CIRCLE & \Circle  & \CIRCLE \\
\hdashline
Security Standards Compliance & \CIRCLE &  \CIRCLE & \CIRCLE & \CIRCLE & \CIRCLE & \CIRCLE & \CIRCLE & \CIRCLE & \Circle &  \CIRCLE\\
\bottomrule
\end{tabular}
\end{table}

In Table~\ref{tab:mapCSaaStoProviders}, the different services described throughout this chapter are mapped to the initially mentioned top ten companies in the managed and professional security services market in Europe according to Frost \& Sullivan. It shows that almost all services are delivered by most of the discussed companies with just a few exceptions. One outstanding exception is cyber insurance. That is because cyber insurance is traditionally provided by traditional insurance companies rather than by tech companies specialising in cyber security services. However, representatives of both sectors do closely collaborate, e.g., regarding consulting and incident response services, as already described in the corresponding section of this chapter. There are even product bundles such as Deutsche Telekom’s “Magenta Security Shield” which includes technical monitoring and response services as well as cyber insurance. Although a bundled offer, the latter is, however, backed by the Allianz insurance company. 

Table~\ref{tab:mapCSaaStoProviders} is based on open-source intelligence, leveraging marketing channels such as the vendor's web sites, service brochures, and white papers which are publicly available via the Internet. If vendors are not mapped to a specific service, it does not necessarily mean that they are not offering this service. Rather, it means that no information regarding this service from the specific vendor could be found at the point in time our investigation took place. Ultimately, what this all means is that the demand for \gls{csaas} and additional security services, will increase in tandem with the expanding threat landscape that has created a real sense of fear across the entire Internet of Everything landscape. 


\clearpage

\printbibliography

\end{document}

%% file: acronyms.tex
\newacronym[plural=CAs,firstplural=certificate authorities (CAs)]{ca}{CA}{certificate authority}
\newacronym{ai}{AI}{artificial intelligence}
\newacronym{arc}{ARC}{Authenticated Received Chain}
\newacronym{bcdr}{BCDR}{Business Continuity / Disaster Recovery}
\newacronym{ceo}{CEO}{Chief executive officer}
\newacronym{ciso}{CISO}{Chief information security officer}
\newacronym{csaas}{CSaaS}{Cybersecurity as a service}
\newacronym{dkim}{DKIM}{DomainKeys Identified Mail}
\newacronym{dmarc}{DMARC}{Domain-based Message Authentication, Reporting and Conformance}
\newacronym{foss}{FOSS}{Free and open-source software}
\newacronym{gdpr}{GDPR}{General Data Protection Regulation}
\newacronym{hr}{HR}{Human resources}
\newacronym{iam}{IAM}{Identity and Access Management}
\newacronym{ics}{ICS}{Industrial Control System}
\newacronym{ioe}{IoE}{Internet of Everything}
\newacronym{iot}{IoT}{Internet of Things}
\newacronym{ir}{IR}{Incident Response}
\newacronym{it}{IT}{Information technology}
\newacronym{mfa}{MFA}{Multi-factor Authentication}
\newacronym{nist}{NIST}{National Institute of Standards and Technology}
\newacronym{ot}{OT}{Operational Technology}
\newacronym{pam}{PAM}{Priviledged Access Management}
\newacronym{pgp}{PGP}{Pretty Good Privacy}
\newacronym{rce}{RCE}{Remote Code Execution}
\newacronym{rpo}{RPO}{Recovery Point Objective}
\newacronym{rto}{RTO}{Recovery Time Objective}
\newacronym{saml}{SAML}{Security Assertion Markup Language}
\newacronym{secaas}{SECaaS}{Security as a Service}
\newacronym{siem}{SIEM}{Security Information and Event Management}
\newacronym{smime}{SMIME}{Secure/Multipurpose Internet Mail Extensions}
\newacronym{spf}{SPF}{Sender Policy Framework}
\newacronym{sso}{SSO}{Single sign-on}
\newacronym{tls}{TLS}{Transport Layer Security}
\newacronym{zta}{ZTA}{zero trust architecture}

%% file: bib.bib
@misc{csa2016,
  author       = {{Cloud Security Alliance - Security as a Service Working Group}},
  title        = {{Defined Categories of Security as a Service}},
  year         = {2016},
  url          = {https://downloads.cloudsecurityalliance.org/assets/research/security-as-a-service/csa-categories-securities-prep.pdf}
}

@misc{isc2022,
  author       = {{International Information System Security Certification Consortium (ISC)²}},
  title        = {{Cybersecurity Workforce Study}},
  year         = {2022},
  url          = {https://www.isc2.org//-/media/ISC2/Research/2022-WorkForce-Study/ISC2-Cybersecurity-Workforce-Study.ashx}
}

@techreport{ThePCinsider2023,
  title={Who Invented the Antivirus? A History of Antivirus Software},
  author={Manish Sahay},
  year={2023},
  institution={thePCinsider}
}

@misc{DeepLocker2018,
  author       = {Kirat, D. and Jang, J. and Stoecklin, M. P.},
  title        = {{DeepLocker - Concealing Targeted Attacks with AI Locksmithing}},
  year         = {2018},
  url          = {https://i.blackhat.com/us-18/Thu-August-9/us-18-Kirat-DeepLocker-Concealing-Targeted-Attacks-with-AI-Locksmithing.pdf}
}

@misc{statista2023,
  author       = {Statista},
  title        = {{Annual number of data compromises and individuals impacted in the United States from 2005 to first half 2022}},
  year         = {2023},
  url          = {https://www.statista.com/statistics/273550/data-breaches-recorded-in-the-united-states-by-number-of-breaches-and-records-exposed/}
}

@misc{embroker2023,
  author       = {Mike Mclean},
  title        = {{2023 Must-Know Cyber Attack Statistics and Trends}},
  year         = {2023},
  url          = {https://www.embroker.com/blog/cyber-attack-statistics/}
}

@misc{technews2023,
  author       = {Troy Beamer},
  title        = {{What Industries Are Most Vulnerable to Cyber Attacks In 2022?}},
  year         = {2023},
  url          = {https://www.techbusinessnews.com.au/what-industries-are-most-vulnerable-to-cyberattacks-in-2022/}
}

@techreport{stephanow2017clouditor,
  title={Clouditor-continuous cloud assurance},
  author={Stephanow, Philipp and Banse, Christian},
  year={2017},
  institution={Fraunhofer AISEC}
}

@misc{frostSullivan2022,
  author       = {{Frost \& Sullivan}},
  title        = {{European Managed and Professional Security Services Market}},
  year         = {2022},
  month        = {8},
  day          = {11},
  url          = {https://store.frost.com/european-managed-and-professional-security-services-market.html}
}

@misc{statista-market-2023,
  author = {{Statista}},
  title = {{Size of the Security as a Service (SECaaS) market worldwide from 2022 to 2033}},
    year         = {2023},
  month        = {1},
  url = {https://www.statista.com/statistics/595164/worldwide-security-as-a-service-market-size/},
}

@misc{IBM-Cybersecurity-2022,
  author = {{IBM Security}},
  title = {{X-Force Threat Intelligence Index 2022}},
    year         = {2022},
  month        = {6},
  note         = "Page 16, Sum of top infection vectors linked to human error.",
  url = {https://www.ibm.com/downloads/cas/ADLMYLAZ},
}

@misc{rfc7522,
    series =    {Request for Comments},
    number =    7522,
    howpublished =  {RFC 7522},
    publisher = {RFC Editor},
    doi =       {10.17487/RFC7522},
    url =       {https://www.rfc-editor.org/info/rfc7522},
        author =    {Brian Campbell and Chuck Mortimore and Michael Jones},
    title =     {{Security Assertion Markup Language (SAML) 2.0 Profile for OAuth 2.0 Client Authentication and Authorization Grants}},
    pagetotal = 15,
    year =      2015,
    month =     may,
}

@misc{rfc6749,
    series =    {Request for Comments},
    number =    6749,
    howpublished =  {RFC 6749},
    publisher = {RFC Editor},
    doi =       {10.17487/RFC6749},
    url =       {https://www.rfc-editor.org/info/rfc6749},
        author =    {Dick Hardt},
    title =     {{The OAuth 2.0 Authorization Framework}},
    pagetotal = 76,
    year =      2012,
    month =     oct,
}

@misc{OpenIDCore,
    url =       {http://openid.net/specs/openid-connect-core-1\_0.html},
        author =    {Sakimura, N. and Bradley, J. and Jones, M. and de Medeiros, B. and C. Mortimore},
    title =     {{OpenID Connect Core 1.0}},
    year =      2014,
    month =     nov
}

@misc{NISTSP800-61Rev2,
  author = {Paul Cichonski and Thomas Millar and Timothy Grance and Karen Scarfone},
  title = {Computer Security Incident Handling Guide},
  year = {2012},
  month = {2012-08-06},
  publisher = {Special Publication (NIST SP), National Institute of Standards and Technology, Gaithersburg, MD},
  doi = {https://doi.org/10.6028/NIST.SP.800-61r2},
  language = {en},
}

@misc{Cybercrime-2019,
  author = {{Robert Johnson}},
  title = {{60 Percent Of Small Companies Close Within 6 Months Of Being Hacked}},
    year         = {2019},
  month        = {1},
  url = {https://cybersecurityventures.com/60-percent-of-small-companies-close-within-6-months-of-being-hacked/},
}

@ARTICLE{9328095,

  author={Khan, Rafiq Ahmad and Khan, Siffat Ullah and Khan, Habib Ullah and Ilyas, Muhammad},

  journal={IEEE Access}, 

  title={Systematic Mapping Study on Security Approaches in Secure Software Engineering}, 

  year={2021},

  volume={9},

  number={},

  pages={19139-19160},

  doi={10.1109/ACCESS.2021.3052311}}

@misc{enisaThreatLandscapeSCA2021,
  title = {{ENISA Threat Landscape for Supply Chain Attacks}},
  author = {{European Union Agency for Network and Information Security (ENISA)}},
  year = {2021},
  url = {https://www.enisa.europa.eu/publications/threat-landscape-for-supply-chain-attacks/@@download/fullReport},
}

@inproceedings{mercat2019,
author = {Heinl, Michael P. and Giehl, Alexander and Wiedermann, Norbert and Plaga, Sven and Kargl, Frank},
title = {{MERCAT: A Metric for the Evaluation and Reconsideration of Certificate Authority Trustworthiness}},
year = {2019},
isbn = {9781450368261},
publisher = {Association for Computing Machinery},
address = {New York, NY, USA},
doi = {10.1145/3338466.3358917},
booktitle = {Proceedings of the 2019 ACM SIGSAC Conference on Cloud Computing Security Workshop},
pages = {1–15},
numpages = {15},
location = {London, United Kingdom},
series = {CCSW'19}
}

@misc{holmsecurity2023,
    url =       {https://www.holmsecurity.com/},
        author =    {Holm Security},
    title =     {Next-Gen Vulnerability Management},
    year =      2023,
    month =     mar
}

@misc{fishing-site,
    url =       {https://www.rmtechteam.com/blog/new-office-365-phishing-tactics-are-difficult-to-spot-but-easy-to-prevent},
        author =    {Holm Security},
    title =     {New Office 365 phishing tactics are difficult to spot but easy to prevent},
    year =      2020,
    month =     jun
}

@misc{Awareness-Market-2023,
  author = {Steve Morgan},
  title = {Security Awareness Training Market To Hit \$10 Billion Annually By 2027},
    year         = {2023},
  month        = {4},
  url = {https://cybersecurityventures.com/security-awareness-training-market-to-hit-10-billion-annually-by-2027/},
}

@misc{Insurance-Market-2023,
  author = {Ankura},
  title = {The Cybersecurity Insurance Market: What to Expect in 2023},
    year         = {2023},
  month        = {3},
  url = {https://www.jdsupra.com/legalnews/the-cybersecurity-insurance-market-what-2446460/},
}
